\documentclass{ieeeaccess}

\usepackage{cite}
\usepackage{array}
\usepackage{url}
\usepackage{amsmath,amssymb,amsfonts}
\usepackage{graphicx}
\usepackage{multirow}
\usepackage{listings}
\usepackage{booktabs}
\usepackage[breaklinks, hidelinks]{hyperref}

\usepackage{algorithmic}
\usepackage[plain]{algorithm}

\usepackage{caption}    
\usepackage{subcaption}
\usepackage{setspace}

\begin{document}

\history{Date of publication xxxx 00, 0000, date of current version xxxx 00, 0000.}
\doi{10.1109/ACCESS.2017.DOI}

\title{The Maximum Common Subgraph Problem: A Portfolio Approach}

\author{
  \uppercase{Andrea Marcelli}\authorrefmark{1},
  \uppercase{Stefano Quer}\authorrefmark{2},
  \uppercase{Giovanni Squillero}\authorrefmark{2} \IEEEmembership{Senior Member, IEEE}
}

\address[1]{Cisco Systems --- Talos Security Intelligence and Research Group, Sophia Antipolis, France}
\address[2]{Politecnico di Torino --- DAUIN, Corso Duca degli Abruzzi 24, 10129 Torino, Italy}

\tfootnote{Authors are listed in alphabetical order.}

\markboth
{Author \headeretal: Preparation of Papers for IEEE TRANSACTIONS and JOURNALS}
{Author \headeretal: Preparation of Papers for IEEE TRANSACTIONS and JOURNALS}

\corresp{Corresponding author: Stefano Quer (e-mail: stefano.quer@polito.it).}

\begin{abstract}
The Maximum Common Subgraph is a computationally challenging problem with countless practical applications. Even if it has been long proven NP-hard, its importance still motivates searching for exact solutions. This work starts by discussing the possibility to extend an existing, very effective branch-and-bound procedure on parallel multi-core and many-core architectures. We analyze a parallel multi-core implementation that exploits a divide-and-conquer approach based on a thread-pool, which does not deteriorate the original algorithmic efficiency and it is not memory bound. We extend the algorithm to parallel many-core GPU architectures adopting the CUDA programming framework, and we show how to handle the heavily workload-unbalance and the massive data dependency. Then, we suggest new heuristics that reorder the adjacency matrix, deal with ``dead-ends'' and randomize the search with automatic restarts, achieving significant improvements on specific cases. Finally, we propose a \textit{portfolio} approach, which integrates all the different local search algorithms as component tools. Such portfolio, rather than choosing the best tool for a given instance up-front, takes the decision on-line. The proposed approach drastically limits memory bandwidth constraints and avoids other typical portfolio fragilities as CPU and GPU versions often show a complementary efficiency and run on separated platforms. Experimental results support the claims and motivate further research to better exploit GPUs in embedded task-intensive, and multi-engine parallel applications.
\end{abstract}


\titlepgskip=-15pt

\maketitle


\section{Introduction}
\label{sec:introduction}

%
%

Graphs are an extremely general and powerful data structure. They can be used to model, analyze and process a huge variety of phenomena and concepts by providing natural machine-readable representations. For instance, graphs may be used to represent the relationships between functions or fragments of code in programs, or the connections among functional blocks in electronic circuits.

%
%

The maximum common subgraph (MCS) problem consists in finding the largest graph which is simultaneously isomorphic to two subgraphs of two given graphs. The problem comes in two forms, namely the \textit{maximum common induced subgraph problem} and the \textit{maximum common partial subgraph problem}. In the former, the goal is to find a graph with as many vertices as possible which is an induced subgraph of each of two input graphs, that is, edges must be mapped to edges and non-edges, to non-edges. In the latter, conversely, the target is to find a common non-induced subgraph, with ``partial'' indicating that only all edges need to be mapped, while some edges might be skipped. In this paper, we discuss the induced variant, and for the sake of simplicity we will refer to it simply as \textit{the} maximum common subgraph problem.

MCS problems arise in many areas, such as bioinformatics, computer vision, code analysis, compilers, model checking, pattern recognition, and have been discussed in literature since the seventies~\cite{barrow1976subgraph,bron1973finding}. MCS has been recognized as NP-hard problem, and it is computationally challenging in the general case: running state-of-the-art algorithms become computationally unfeasible on problems with as little as $40$ vertices, unless some meta information allows to reduce the search space. Practitioners from many fields are still asking researchers how to effectively tackle the problem in their real-world applications.

%
%

In 2017 McCreesh et al.~\cite{mccreesh2017partitioning} proposed McSplit, a very efficient branch-and-bound algorithm to find maximum common subgraphs for a large variety of graphs, such as, undirected, directed, labeled with several labeling strategies, etc. Essentially, McSplit is a recursive procedure adopting a smart invariant and an effective bound prediction formula. The invariant considers a new vertex pair, within the current mapping between the two graphs, only if the vertices within the pair share the same \textit{label}, i.e., they are connected in the same way to all previously selected vertices. The bound consists in computing, from the current mapping size and from the labels of yet to map vertices, the maximum MCS size the current recursion path may lead to, and to prune those paths that are not promising enough.

In this paper, we extend McSplit in two directions. First, we present a multi-core and a many-core parallel implementations. In the former, CPU-based version, threads are organized in a thread pool, sharing a queue in which \textit{tasks} are inserted by the the working thread and later extracted and solved by the first free helper thread. Mechanisms are used to avoid task duplication, task unbalance, and data structure repetition, as such problems may impair the original divide-and-conquer approach. Then, we analyze how to extend the approach to a many-core GPU, adopting the CUDA programming framework. We describe the corrections required to handle the heavy workload unbalance and the massive data dependency of the original, recursive algorithm.

After that, we extend the classical ``winner-takes-all'' approach to the MCS problem to a multi-engine methodology~\cite{SATcomp,SMTcomp,10.1007/978-3-319-50349-3_8}. However, the structural properties of graphs not always relate with the difficulties faced by the MCS, and defining a representative set of problem instances to assess tools' performances is not feasible. Unlikely standard multi-engine approaches, in the proposed solution the different algorithms run in parallel, and the algorithm selection is made on-line.

We insert in our portfolio the original sequential and parallel versions, improved with new heuristics. Namely, we present variations to compute the original order in which vertices are paired, we suggest counter-measures to deal with the ``heavy-tail phenomenon'', and we propose randomize tree search and automatic search restarts. The resulting algorithms offer very good performance on specific instances and are thus perfect candidates to belong to a portfolio. Moreover, as they resort to different computation units, CPU and GPU cores, both the overall memory bandwidth and computation power are increased, reducing the typical multi-engine fragilities.

Experiments were ran on standard benchmarks, coming from~\cite{DeSanto2003,foggia2001database}. We compare the original McSplit implementations against the many-core, multi-core and portfolio implementations. We also consider some intermediate versions, generated to move from the multi-core to the many-core and to the multi-engine approach, detailing advantages and disadvantages of each version.

%
%

The paper is organized as follow: Section~\ref{sec:background} reports some background on graph isomorphism, and it summarizes McSplit; Sections~\ref{sec:cpu},~\ref{sec:gpu} and~\ref{sec:portfolio} introduce our parallel multi-core, many-core, and multi-engine (portfolio) implementations, respectively; Section~\ref{sec:expRes} reports our evidence analysis; Section~\ref{sec:conclusion} includes some final considerations.


\section{Background}
\label{sec:background}


\subsection{The Maximum Common Subgraph Problem}

Finding the similarity of two graphs has been the subject of extensive research in the last decades~\cite{bunke2002comparison,conte2007challenging,vismara2008finding,minot2014searching,chen2015approximating,mccreesh2017partitioning,Bunke:2003:GCU:1757868.1757895,doi:10.1137/S0036144502415960,zagerThesis}.

\begin{figure}[ht!]
\centering
\begin{subfigure}[b]{0.10\textwidth}
  \centering
  \includegraphics[width=1.0\textwidth]{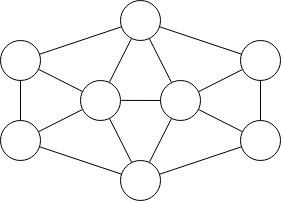}
  \caption{
    \label{fig:graph_a}
  }
\end{subfigure}
\hfill
\begin{subfigure}[b]{0.10\textwidth}
  \centering
  \includegraphics[width=1.0\textwidth]{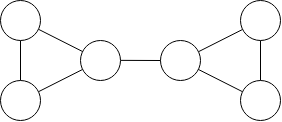}
  \caption{
    \label{fig:graph_b}
  }
\end{subfigure}
\hfill
\begin{subfigure}[b]{0.10\textwidth}
  \centering
  \includegraphics[width=1.0\textwidth]{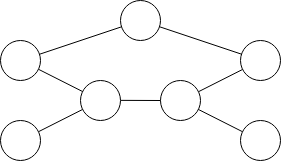} 
  \caption{
    \label{fig:graph_c}
  }
\end{subfigure}
\hfill
\begin{subfigure}[b]{0.10\textwidth}
  \centering
  \includegraphics[width=1.0\textwidth]{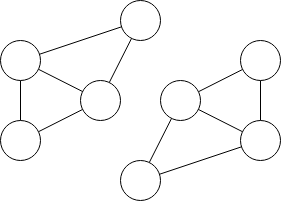} 
  \caption{
    \label{fig:graph_d}
  }
\end{subfigure}
\caption{
  \label{fig:graphs_samples}
  An undirected graph with induced and non-induced subgraphs.
  Fig.~(a): The original undirected graph with 8 vertices and 15
  edges.
  Fig.~(b): An induced subgraph.
  Fig.~(c): A non-induced subgraph.
  Fig.~(d): A non-connected subgraph.
}
\end{figure}

Given a graph $G = (V_G, E_G)$, a graph $H = (V_H, E_H)$ is a \textit{subgraph} of graph $G$ (i.e., $H \subseteq G$) if it is composed by a set of vertices such that $V_H \subseteq V_G$. If $H$ includes all the edges $e \in E_G$ with both the endpoints in $V_H$, it is called an \textit{induced} subgraph. Otherwise, $H$ is called a \textit{non-induced} or \textit{partial} subgraph. Fig.~\ref{fig:graphs_samples} shows some examples of induced and of non-induced subgraphs. Given two graphs, $G$ and $H$, the graph $S$ is a \textit{common subgraph} if $S$ is simultaneously isomorphic to a subgraph of $G$ and to a subgraph of $H$. 

\begin{figure}[ht!]
\centering
\begin{subfigure}[b]{0.49\textwidth}
  \centering
  \includegraphics[width=0.55\textwidth]{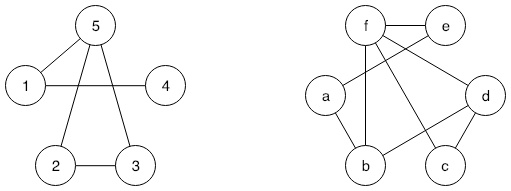}	
  \caption{
    \label{fig:mcs_a}
    Undirected graphs $G$ and $H$.
  }
\end{subfigure}
\hfill
\begin{subfigure}[b]{0.49\textwidth}
  \centering
  \includegraphics[width=0.9\textwidth]{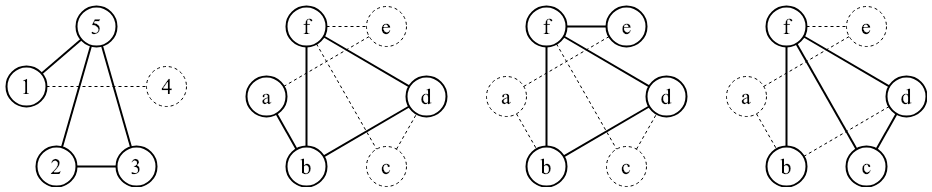}	
  \caption{
    \label{fig:mcs_b}
    Different MCS mappings for $G$ and $H$.
  }
\end{subfigure}
\caption{
  \label{fig:mcs}
  Given the graphs $G$ and $H$ represented in Fig.~(a),
  Fig.~(b) reports 4 different possible MCSs.
  Labels are shown only to help identify the vertices.
}
\end{figure}

A MCS of $G$ and $H$ is the largest possible common induced subgraph, i.e., the common subgraph with as many vertices as possible. Fig.~\ref{fig:mcs} shows some examples with undirected and unlabeled graphs.

In the rest of the paper, we will indicate with $|V_G|$ and $|V_H|$ the number of vertices of $G$ and $H$, respectively.

\subsection{The McSplit Algorithm}
\label{sec:mcSplit}

In 2017 McCreesh et al.~\cite{mccreesh2017partitioning} proposed McSplit, so far one of the most effective algorithm to find maximum common subgraphs. Different implementations of the algorithm are provided by the authors themselves, but essentially it is a recursive procedure based on two main ingredients: A smart invariant and an effective bound prediction formula.

\begin{figure}[ht!]
\centering
\begin{subfigure}[b]{0.10\textwidth}
\includegraphics[width=1.0\textwidth]{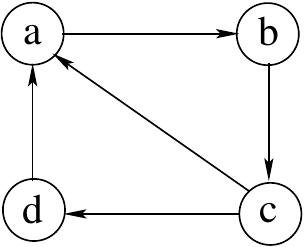}
\caption{
  \label{fig:mcsExa}
}
\end{subfigure}
\hfill
\begin{subfigure}[b]{0.10\textwidth}
\includegraphics[width=1.0\textwidth]{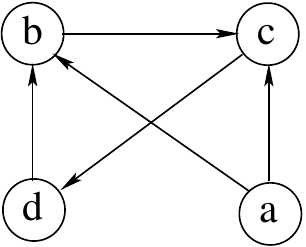}
\caption{
  \label{fig:mcsExb}
}
\end{subfigure}
\hfill
\begin{subfigure}[b]{0.10\textwidth}
\includegraphics[width=1.0\textwidth]{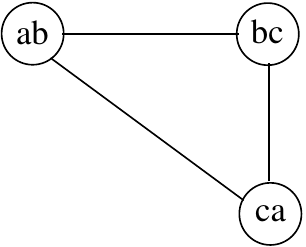}
\caption{
  \label{fig:mcsExc}
}
\end{subfigure}
\hfill
\begin{subfigure}[b]{0.10\textwidth}
\includegraphics[width=1.0\textwidth]{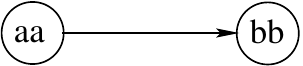}
\caption{
  \label{fig:mcsExd}
}
\end{subfigure}
\caption{
  \label{fig:mcsEx}
  A running example.
  The graphs $G$ and $H$, of Fig.~(a) and Fig.~(b), produce
  the MCS of Fig.~(c) if they are considered undirected and unlabeled,
  and the MCS of Fig.~(d) if they are considered as directed and labeled.
}
\end{figure}

The procedure dovetails together the two ingredients as follow. Let us suppose we have the two graphs $G$ and $H$ represented in Fig.~\ref{fig:mcsExa} and~\ref{fig:mcsExb}. We consider graphs as simple as possible for the sake of readability. For now, we also suppose these graphs are undirected (even if they are represented as directed in Fig.~\ref{fig:mcsEx}) and unlabeled, i.e., the nodes labels $\{a, b, c, d\}$ are reported only to uniquely identify all vertices.

McSplit builds a mapping $M$ between the vertices of $G$ and of
$H$ using a depth-first search.
At the very beginning $M$ is an empty set, and the algorithm adds a
vertex pair to $M$ at each recursion level.
During the first recursion step, let us suppose we arbitrarily select
vertex $a$ in $G$, and we try to map this vertex with vertex $b$ in
$H$, that is $M = \{ a, b \}$.
Now we label each unmatched vertex in $G$ according to whether it is
adjacent to vertex $a$, and we label each unmatched vertex in $H$
according to whether it is adjacent to vertex $b$.
For undirected graphs, adjacent vertex have label 1 and non-adjacent
vertex have label 0.
Table~\ref{tab:label}(a) shows these labels.
At this point, we recur and we try to extend $M$ with a new pair.
A new pair can be inserted into $M$ only if the vertices within the
pair share the same label.
This is McSplit main invariant.
If, during the second recursion step, we extend $M$ with pair $b$ and
$c$, we obtain the new mapping $M = \{ ab, bc \}$, and the new label
set represented in Table~\ref{tab:label}(b).
The third step consists of extending $M$ with $c$ and $a$, which is,
at this point, the only remaining possibility.
We obtain $M = \{ abc, bca \}$ and the label set of
Table~\ref{tab:label}(c).
The last two vertex $d$ and $d$ cannot be inserted int $M$ because
they have different labels.
Then the recursive procedure backtracks and it looks-for another
(possibly longer) match $M$.
After all possibilities have been exhaustively explored, the final
result, i.e., the MCS of $G$ and $H$ is represented in
Fig.~\ref{fig:mcsExc}.

\begin{table*}[htbp]
\small
\begin{center}
\begin{tabular}{|cc|cc|c|cc|cc|c|cc|cc|}
\multicolumn{2}{c}{$G$} & \multicolumn{2}{c}{$H$} &
\multicolumn{1}{c}{} &
\multicolumn{2}{c}{$G$} & \multicolumn{2}{c}{$H$} &
\multicolumn{1}{c}{} &
\multicolumn{2}{c}{$G$} & \multicolumn{2}{c}{$H$} \\
\cline{1-4}
\cline{6-9}
\cline{11-14}
\multicolumn{1}{|c}{v} &
\multicolumn{1}{c}{Label} &
\multicolumn{1}{|c}{v} &
\multicolumn{1}{c|}{Label} &
\multicolumn{1}{|c}{~~~~~~~~~~} &
\multicolumn{1}{|c}{v} &
\multicolumn{1}{c}{Label} &
\multicolumn{1}{|c}{v} &
\multicolumn{1}{c|}{Label} &
\multicolumn{1}{|c}{~~~~~~~~~~} &
\multicolumn{1}{|c}{v} &
\multicolumn{1}{c}{Label} &
\multicolumn{1}{|c}{v} &
\multicolumn{1}{c|}{Label} \\
\cline{1-4}
\cline{6-9}
\cline{11-14}
b & 1 & c & 1 & & c & 11 & a & 11 & & d & 101 & d & 110 \\
\cline{11-14}
c & 1 & a & 1 & & d & 10 & d & 11 \\
\cline{6-9}
d & 1 & d & 1 \\
\cline{1-4}
\end{tabular}
\caption{
  \label{tab:label}
  Labels on the non-mapped vertices of $G$ and $H$ with mapping:
  Fig.~(a) $M = \{ a, b \}$,
  Fig.~(b) $M = \{ ab, bc \}$. and
  Fig.~(c) $M = \{ abc, bca \}$.
}
\end{center}
\end{table*}

As previously mentioned, the second very important ingredient of the
algorithm is the bound computation.
This computation is used to effectively prune the space search.
While parsing a branch by means of a recursive call, the following
bound is evaluated:
\begin{equation}
\label{eq:bound}
\small
\begin{array}{ccl}
bound & = &
|M| + \sum_{l \in L} min(| \{ v \in G : label(v)=l \} |, \\
& & \hspace*{18mm} | \{ v \in H: label(v)=l \} |)
\end{array}
\end{equation}
where $|M|$ is the cardinality of the current mapping and $L$ is the
actual set of labels.
If the bound is smaller than the size of the current mapping, there is
no reason to follow that path along the tree search, as there is no
possibility to find a match longer than the current one in there.
In this way the algorithm prunes consistent branches of the decision
tree, drastically reducing the computation effort.

As the MCS problem comes in many variants, McCreesh et al. propose small algorithmic variations to deal optimally with directed, vertex-labeled and edge-labeled graphs. For example, if we consider the graph in Fig.~\ref{fig:mcsExa} and~\ref{fig:mcsExb} as directed and with node labels, and we apply the previous process incrementing it with the two previous modifications, we find the MCS reported in Fig.~\ref{fig:mcsExd}.

\section{Parallel Approaches}
\label{sec:cpuAndgpu}

In this section we present our parallel approaches, the CPU-based multi-core in Section~\ref{sec:cpu}, and the GPU-based many-core in Section~\ref{sec:gpu}.


\subsection{The CPU multi-core Approach}
\label{sec:cpu}

While Trimble himself briefly discusses a parallel multi-thread
version to efficiently parallelize the algorithm analyzed in
Section~\ref{sec:mcSplit}~\cite{trimble2017parallelCPP,mccreesh2017PhD,hoffmann2018observations},
no complete description of the concurrent MCS function is available in
the literature.
 
\begin{figure}[ht]
  \centering
  \includegraphics[width=0.35\textwidth]{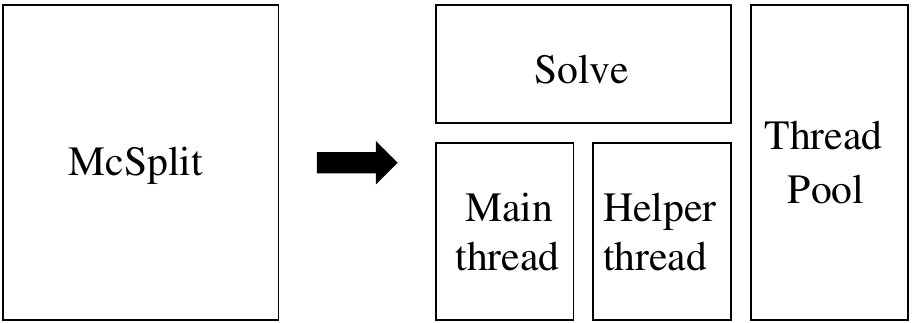}
  \caption{
    \label{fig:McSplitSplitted}
    Partitioning the McSplit function to parallelize it.
  }
\end{figure}

To implement the concurrent version of the algorithm, the original recursive procedure McSplit has been decomposed and restructured to balance the workload among the threads. The core delegates to a set of helper threads portions of the job done by the original function during its deeper recursion levels (fig.~\ref{fig:McSplitSplitted}).

The top-level function {\sc Solve} orchestrates all worker threads. It initializes the matching process among vertices, and in order to achieve independence between working threads, it performs a copy of all relevant data structures, before moving into recursion. Depending on the recursion level, function {\sc Solve} goes on calling a \textit{main thread} or an \textit{helper thread}. 

The concurrency level has been tuned not to waste CPU-time copying data, and not overload the task queue in the thread-pool with too short tasks, avoiding slowdowns of the algorithm. During the first few levels of the recursion, each iteration of the main loop that pairs vertices from the two graphs is very time consuming; Then, the deeper in the recursion tree, the shorter become each branch. We define a threshold for the maximum depth in the recursion tree above which the function stops delegating work to other threads. Only main threads are called after the threshold. That is, delegation is not used when splitting the current task among more threads would be too expensive and would probably generate too many working threads. 

Helper threads are called at lower recursion levels to help the main thread to solve its current task. It must be remarked that the \textit{main thread} is not the first thread started by the operating system, and that in different phases of the algorithm, any thread can act as a main thread. Each thread executing the main function of a particular branch of the recursion, and eventually asking for help, acts as a \textit{main thread}. At the same time, each thread helping a main thread acts as an \textit{helper thread}.

In this scenario, threads cooperate with each other to compute, in parallel, different iterations of the main cycle in which pair of vertices are coupled depending on their labels. Thus the main task of all \textit{main threads} and \textit{helper threads} is to go on suggesting new matching pairs, and eventually recurring on their own recursion path. The main difference between \textit{main} and \textit{helper threads} is that the latter make a copy of all incoming data structure before moving on. The main and the helper functions are designed such that they are data-independent, and they can be assigned to different threads to be run in parallel. Anyway, they are actually linked by two atomic variables. The first one is used as a shared index of the main iterative construct. In this way, the main and the helper thread can execute the same cycle, but they must distribute iterations without both executing the same ones. The second atomic variable is the size of the current matching, which is shared among all threads to correctly evaluate the opportunity to proceed along a new path.

To further increase efficiency, all threads are kept running during the entire execution of the algorithm, and they are orchestrated using a \textit{thread pool}. Such a pool implements a priority queue in which threads can insert the tasks they want to be helped with. These tasks include copies of the data structures and a pointer to the helper function. Helper threads can pick-up new tasks from the queue as soon as they are available. To achieve a higher level of parallelism, more than one helper thread can pick the same task and cooperate to solve it, also if at different times. The task objects are put in the queue following the order given by their position in the recursion three. For example, a task generated at depth $d$ will be put in the queue before any tasks generated at depth $d+1$. In the same way, a task generated at depth $d$ in the iteration $i_n$ of the main loop will be put after a task generated at depth $d$ in the first iteration $i_0$. The shared variables avoid repeating the same tasks more than once.


A simplify version of this entire process is represented by
Algorithm~\ref{alg:mcsParallel}.
The top level procedure {\sc Parallel\_McSplit} receives
the original graphs $G$ and $H$.
It sets the current mapping $M$, and the best mapping
$M_{best}$, to the empty set, and it initializes the priory queue
$Q$ used by the thread-pool.
After that, it calls function {\sc Solve} which is essentially a
recursive function adopting the concurrent scheme previously
described.
{\sc Solve} receives as parameters the two graphs $G$ and
$H$, the current and the best mapping $M$ and $M_{best}$, and the
recursion level depth (properly initialize to 0).
At each run of the recursive function, if a greater mapping has been
found (lines 6--8). the best incumbent graph $M_{best}$ is updated.
Then, the upper bound for the current search branch is computed (line
9) using Equation~\ref{eq:bound}.
If the bound has been hit (i.e., the current bound is lower or equal
to the size of actual incumbent, thus no future search can improve
it), the branch is pruned and the function returns (line 11).
Otherwise, a new correct mapping is searched.
In this case, first the most promising label class is selected from
the remaining set of label classes (line 13), according to some
heuristic.
Then, a vertex belonging to that label class is selected as a new
vertex $v$ (line 14).

Finally, the iterative cycle of lines 15--25 visits every vertex
$u$ of $H$ belonging to the previously selected class.
For each of these vertices, the recursive function explores the
consequences to add the pair $(v, u)$, to the mapping (line 16).
Notice that considering $(v, u)$ as a new pair in the mapping, implies
splitting the current classes into two new sub-classes, according to
whether vertices belonging to it are adjacent or not to $v$ and $u$.
This operation if performed by function {\sc filterDomain} at lines 17
and 18.
At this point, as previously described, we resort to helper threads for
the lower recursion levels (lines 20--21) and we proceed independently
for higher depths (line 23).
In the first case, a new pool task is first generated (line 20) and
then enqueue in the thread pool queue (line 21).
In the second case, a new recursive call is performed with updated
$G$, $H$, $M$, $M_{best}$, and the recursion depth (line 23).

Notice that the process reported in lines 19--24 is repeated in lines
27--32.
This last section of the code is required to check the case in which
the vertex $v \in G$ is not mapped at all with any vertex of $H$.
Thus $v$ is ruled-out from the graph $G$ (line 26) and the process
proceeds either concurrently or recursively.

\begin{algorithm}[!htb]
\centering
\renewcommand{\algorithmicrequire}{\sc Parallel\_McSplit
  ($G$, $H$)}
\begin{minipage}{25pc}
\begin{algorithmic}[1]
\REQUIRE
\STATE $M$ = $\emptyset$
\STATE $M_{best}$ = $\emptyset$
\STATE {\sc QueueInit} (Q)
\STATE {\sc Solve} ($G$, $H$, $M$, $M_{best}$, 0)
\STATE {\sc QueueDestroy} (Q)
\\~
\renewcommand{\algorithmicrequire}{\sc Solve
  ($G$, $H$, $M$, $M_{best}$, depth)}
\REQUIRE
\IF {($|M| > |M_{best}|$)} 
\STATE $M_{best}$ = $M$
\ENDIF
\STATE bound = {\sc computeBound} ($G$, $H$, $M$)  
\IF {(bound $\leq$ $|M_{best}|$)} 
\STATE  return
\ENDIF
\STATE bidomain = {\sc selectLabelClass} ($G$, $H$)
\STATE $v$ = {\sc selectVertex} (bidomain)
\FORALL{$u \in H$}
\STATE $M = M \cup (v, u)$
\STATE $\widehat{G}$ = {\sc filterDomain} ($G$, $v$)
\STATE $\widehat{H}$ = {\sc filterDomain} ($H$, $u$)
\IF {(depth $\leq$ PART\_LEVEL)}
\STATE task = \{ $\widehat{G}$, $\widehat{H}$, $M$,
  $M_{best}$, depth \}
\STATE {\sc Enqueue} (Q, {\sc Solve}, task)
\ELSE
\STATE {\sc Solve} ($\widehat{G}$, $\widehat{H}$, $M$, $M_{best}$, depth+1)
\ENDIF
\ENDFOR
\STATE $G = G\backslash v$
\IF {(depth $\leq$ PART\_LEVEL)}
\STATE task = \{ $G$, $H$, $M$, $M_{best}$, depth \}
\STATE {\sc Enqueue} (Q, {\sc Solve}, task)
\ELSE
\STATE {\sc Solve} ($G$, $H$, $M$, $M_{best}$)
\ENDIF
\STATE return
\end{algorithmic}
\end{minipage}
\caption{
  \label{alg:mcsParallel}
  The multi-core recursive branch-and-bound McSplit procedure.
}
\end{algorithm}

As a last comment, notice that a fundamental role in the system is
played by the queue within the thread pool and by the working
threads running for the entire process and waiting for new tasks in
the queue.
In fact, the thread pool guarantees a very balanced work-load among
threads: Any time a thread finishes a branch, it can go through the
tasks queue and start helping other threads with their branches, thus
reducing its idle time.
This mechanism is not illustrated by the pseudo-code for the sake of
simplicity, but depending of the status of the queue, various
execution flow can be executed by threads:
\begin{itemize}
\item
The task queue is empty.
When the program starts and the thread pool is first initialized, each
thread will start its own cycle when the task queue is empty.
In this situation each thread acquires the lock on the queue (not
everyone at the same time, of course), and then it tries to pick a
task from the queue.
Since the latter is empty, the thread skips all operations without
doing anything, and the thread blocks on the condition variable
waiting for someone to put a task in the queue and wake it
up\footnote{POSIX mutex and condition variable objects are used but
not detailed in here}.
\item
Every time the {\sc Solve} function adds a task to the queue, each
thread of the thread pool that in that moment is waiting for the
condition variable to become signalled is released.
The first thread which manages to acquire the lock proceeds within
its main cycle, and it picks the task from the queue.
Even it this task should no longer be executed by any other thread,
the only entity allowed to remove it from the queue is the one that
put it in the queue.
\item
The task queue contains already executed tasks.
The tasks queue is ordered by the position order of the tasks
themselves, and thus tasks are executed and completed following the
order of the recursion.
Any completed task that is still being executed by some thread, will
be at the beginning of the queue.
When a task is executed it will be marked as such (if will have a
{\tt NULL} pointer instead of the actual pointer to the function which
must solve the problem), such tasks will be simply skipped by any
thread scanning the queue searching for a task to execute.
When a task with a non-NULL pointer is found, the execution continues
as the previous item point.
If no task is found, the execution continues as in the first point. 
\end{itemize}
%


\subsection{The GPU Many-core Approach}
\label{sec:gpu}

In this section, we show how the algorithm described in Section~\ref{sec:cpu} can be modified in order to make it suitable to run on NVIDIA GPUs. We will concentrate on the most effective strategies and optimizations among the several different ones that have been implemented.

As shown in previous sections, McSplit is intrinsically recursive and the maximum recursion depth reached is equal to the size of the final solution graph. Unfortunately, even using Dynamic Parallelism, CUDA does not support more than $24$ in-depth recursion levels, and it is then suitable for graphs with a maximum of $24$ nodes. For this reason, the first step necessary to adapt the algorithm to CUDA, is to modify the execution structure in order to formally remove recursion. Substituting recursion explicitly using a stack is practically unfeasible on GPUs: First of all, as the stack would store every arguments for each virtual recursion, this approach would use a lot of memory and in an unpredictable way, whereas the memory for each thread memory on a GPU is quite limited. Moreover, each stack frame would correspond to a branch of the algorithm with its own varying depth, and the approach would lead to an extremely unbalanced workload on each GPU thread. In the next subsections, we analyze a few possibilities to decrease the algorithm memory usage and increase work-balance, while maintaining the absence of recursion.


\subsubsection{Labels re-computation}

A first attempt is to get rid of the bidomain data structures and to drastically reduce the amount of information stored into the stack, is the following one. Each new stack entry consists of a pair of integers values. Each pair indicates a possible mapping between vertices of the two graphs. The current position in the final solution is stored into a proper variable, whose value is increased and decreased using two different special (and reserved) entries in the stack.

For example, let us suppose we are searching the MCS for two graphs
($G$ and $H$) with 5 vertices.
At the beginning of the algorithm, any combination of one vertex from
the first graph and one from the second graph is possible, and the
stack will contain all those pairs, from
$\langle 4,4 \rangle$ down to
$\langle 0,0 \rangle$
(in reverse order, so they are popped out from the stack in the same
order than the standard algorithm).
When we pop the first pair, namely $\langle 0,0 \rangle$, we
also compute the new label classes for each remaining vertex
from both the input graphs.
Then, we must push in the stack all combinations of those pairs
composed by vertices with matching labels.
Anyhow, before pushing such pairs on the stack we push the pair
$\langle -1,0 \rangle$, and after pushing them we also insert the pair
$\langle 0,-1 \rangle$.
In this way, when we pop the pair
$\langle 0,-1 \rangle$
we know that the current solution position is increased from 0 to 1.
When we do that the algorithm starts popping pairs that it has to
insert in the second position of the solution.
At the same time, when the pair
$\langle -1,0 \rangle$
is extracted, the solution position returns to 0.
Obviously, every time a better solution is found, it is stored in the
incumbent solution, similarly to the standard algorithm.


Unfortunately, a fundamental aspect of this solution is that it
does not allow to efficiently calculate the bound for a new branch.
Indeed, every time a new pair is selected and inserted into the
current solution, we must re-evaluate the labels for each remaining
vertex to count the number of vertices with the same label we have, to
finally compute the bound with Equation~\ref{eq:bound}.
Moreover, the necessity to store separately each vertex pair belonging
to the same label class, instead of grouping them into a bidomain that
is able to contain them all, leads to stack whose size is quickly
increasing.
These two drawbacks lead to massive slowdowns and make questionable
its utilization in order to solve the problem in reasonable time.


\subsubsection{New bidomains stack}

To recover from the drawbacks highlighted in the previous sub-section,
our target is to group vertices in label-classes but keeping the stack
small enough and maintaining the algorithm iterative.

To reach this target, we redesigned the stack to store bidomains data
in an optimized form.
New items are pushed into the stack every time a new set of bidomains
is created, but they are popped out only after every pairs of vertices
contained in the bidomains has been selected and explored.
The stack thus become a matrix with eight columns, and a variable number
of rows.
Each column stores numbers representation at most the degree of the
graph, thus we decided to use unsigned characters limiting the graphs
size to $254$ vertices (the value $255$ is reserved by the algorithm).
In this way each stack entry is four byte long, as in the previous
implementation, in which each entry contained two integer values.
Machines able to deal with graphs larger than $254$ vertices should
also provide much more memory, so the same algorithm could be adjusted
to work with unsigned integers too.
Moreover, stack entries of eight bytes allow better memory alignment
of the stack that can lead to benefits in memory performances and
addresses calculation.

The first five values actually correspond to the ones included in the
bidomain data structure with the same meaning (namely, left, right,
left length, right length, and adjacent).
The remaining three columns are used to store values that are
necessary to restore values of certain variables when backtracking, as
we need to explicitly store some values from previous iterations in
the stack and we cannot resort to local variables within the recursive
procedure.

Similarly, the current solution and the incumbent structure have been
replaced by simple matrices of two rows and a variable number of
columns, storing again unsigned characters.

Given this data structure, similarly to the standard algorithm, at
each iteration of the main cycle, we choose a vertex from left
bidomain and a vertex from right bidomain, then we put them in the
solution and we compute the new bidomains depending on which vertices
we just chose.
However, in the recursive algorithm the selection of a vertex is
performed by swapping it with the one at the end of the domain and by
decreasing the bidomain size.
In the recursive algorithm, this is possible because when backtracking
to higher recursion levels, deep bidomains are destroyed, and the ones
at higher levels has not been touched by lower recursions, thus the
algorithm can continue with consistent data structures.
When recursion is removed, however, we loose this important property,
and when we select a vertex from the right bidomain and we decrease its
size, the original size must be stored somewhere in order to be able
to restore it when every right vertices has been tried with the
selected left vertex.
So we can select another left vertex and try again to pair it with
every right vertex in the same bidomain.
The values of the last column of the stack have precisely this
purpose, every time a new bidomain is creates, is there stored the
initial value of \texttt{RL}, for future utility.

In a similar way, we also need to recreate the behavior of the
\texttt{for} loop of the recursive \texttt{solve} function, but since
we have not a recursive call for each iteration, and the whole
\texttt{while} main cycle is re-executed every time, we introduced an
element in the stack that keep track every time of which right vertex
has been selected, in order to correctly pick the next one.

Following this approach, every time a bidomain has been completely
explored, we remove it from the stack and we pick the immediately
following one.
Doing so, we lose the possibility to choose which bidomain to select
at each time.
Indeed, such an approach, without any further optimization,
performs significantly worse than the original on graphs larger than
15 nodes.
However, it is pretty easy to implement a function to select a new
label class similar to the original one.
The idea, in this case, is to visit the stack backward (from top to
bottom) to select the new bidomain following the desired
heuristic and eventually exchange it with the one in the top of the
stack.


\section{The Portfolio Approach}
\label{sec:portfolio}

The so-called ``portfolio approaches'' have been proposed and adopted in several domains. In such approaches, researchers drop the idea of designing a single, optimal algorithm, and resort to a \textit{portfolio} of different \textit{tools}, each one able to tackle specific instances of a given problem. The relationship between the properties of a problem instance and tool's performance is typically opaque and difficult to capture, the construction of a portfolio normally involves machine-learning activity.

Portfolio approaches were demonstrated the best approach for model checking and satisfiability in recent competition contexts~\cite{jsat2014,SATcomp,SMTcomp}. In these scenarios, portfolio approaches have often been used in order to emphasize multi-core CPU exploitation, and to encourage multi-threaded solutions~\cite{Bordeaux2003ExperimentsWM}.  Moreover, portfolio approaches have been proved to be more diversified, and less sensitive to model characteristics~\cite{SATzilla2007,Pulina2009,Hamadi09manysat}.

Generally speaking, portfolio tools take as input a distribution of problem instances and a set of tools, and constructs a portfolio, optimizing a given objective function. The function can be related to the mean runtime, percent of instances solved, or score in a competition. In our case, we will focus on the number of instances solved in a specific slotted wall-clock time\footnote{The wall-clock time is the difference between the time at which the task finishes and the time at which the task started. It is also known simply as \textit{elapsed time}. }. Using wall-clock time versus using only process time, both as time limit and for tie-breaking, has the advantage to encourage parallelization of tools. Indeed, the concurrency level of each tool can often be guessed only from experimental data, comparing CPU and wall-clock run times. Furthermore, in our portfolio, we may not only have different solvers possibly running in parallel on the CPU cores, but at least one version running on the GPU, thus augmenting the range of variations and affecting different hardware devices, with separate resources.

We built our portfolio starting with the following code versions:
\begin{itemize}
\item
Versions ($mc1$ and $mc2$), are the original versions (sequential
and parallel, respectively) implemented by McCreesh et
al.~\cite{mccreesh2017partitioning}, in C++.
\item
Version 1 ($v1$) is a sequential re-implementation of the original
code in C language.
This code constitutes the starting version for all our
implementations, as it is more coherent with all following
requirements, modifications, and choices.
\item
Version 2 ($v2$) is the C multi-thread version described in
Section~\ref{sec:cpu}.
It logically derives from $mc1$ and $v1$.
\item
Version 3 ($v3$) is an intermediate CPU single-thread implementation
that removes recursion and decreases memory usage.
It is logically the starting point for comparison for the following
two versions.
\item
Version 4 ($v4$) is a CPU multi-thread implementation based on the
same principles of the following CUDA implementation.
\item
Version 5 ($v5$) is the GPU many-thread implementation, described in
Section~\ref{sec:gpu}.
It is based on $v3$ and $v4$.
\end{itemize}
In the sequel, we will refer to the portfolio tool itself with $vP$.
Notice that while the original version $mc1$ and $mc2$ were written in
C++, all our versions have been written in C language but $v5$ which
is written in CUDA.
Note that versions $v3$ and $v4$ have been developed to explore
the feasibility of new approaches in developing a maximum common
subgraph algorithm in CUDA environment, not to obtain the best
possible performances.
For example, $v4$ is drastically slower than all other versions, but
it represents the possibility to run the algorithm implemented in
version 3 on more threads, which are completely independent from one
another.
Indeed the same system is used on a wider scale in version 5, which is
significantly faster than $v4$, although they make use of the same
approach.

We also consider the clique encodings of McCreesh et
al.~\cite{mccreesh2016clique}, and the k$\downarrow$ algorithm of Hoffmann
et al.~\cite{hoffmann2017clique}.
To increase the range of variations we also modified versions $v2$,
$v3$, and $v5$ as described in the following section.


\subsection{Portfolio-based Modifications}
\label{sec:portfolioModification}


An efficient branch-and-bound procedure requires a good initial
ordering to rearrange the recursion tree and improve the convergence
speed.
We propose an approach based on the analysis and permutation of the
adjacency matrix of the two graphs to produce good recursion trees,
which may significantly affect the time and memory requirements of the
computation.
When ordering an adjacency matrix for graph isomorphism, the rows of
the matrix correspond to a permutation of the initial pairing scheme.
We assume that the order is top-down.
A straightforward analysis suggests that it is advantageous to pair
variables with similar support sets.
Mc Creesh et al.~\cite{mccreesh2017partitioning} always sort the
vertices in descending order of degree (or total degree, in the case
of directed graphs).
Anyhow, as the same authors noticed, some improvements is possible in
that direction.

In mathematics, a block matrix is a matrix that can be partitioned
into a collection of smaller matrices each one including a block
of horizontal and vertical lines.
If we consider the adjacency matrix of a directed graph, the presence
of more than one connected component in the graph signals the
existence of non-trivial block-diagonal form for the matrix.
Identifying the connected components of the graph and then ordering
each component individually is desirable.
Finding the connected components is linear in the size of the matrix,
even for full matrices, and leads to decomposition of the problem.
The overall order for a matrix that has multiple connected components
is obtained by concatenating the orders for each component.
The sorting of the components is based on the ratio of number of
rows to number of columns.
If the ratio is large, the component goes toward the beginning of the
schedule.

Within each component, we order the vertices based on their bit
relations.
A square matrix is called upper triangular if all the entries below
the main diagonal are zeros.
Our ordering procedure is roughly based on the algorithm of Hellerman
and Rarick~\cite{1972Hellerman} to put a matrix in bordered block
lower-triangular form.
Initially, all non-zero rows and columns of the matrix block form the
active submatrix.
The procedure then iteratively removes rows and columns from
the active block submatrix and assigns them to final positions.
It iteratively chooses the column that intersects the maximum number of
shortest rows of the active submatrix.
Then, it moves that column out of the active block submatrix and
immediately to its left.
Once the active block submatrix vanishes, the block matrix has been
permuted.


As analyzes in Section~\ref{sec:mcSplit} the recursive procedure
adopts a quite effective bound prediction formula to constrain
the tree search.
However, this constraining effect is reduced when the value of the
current bound gets closer to the actual one.
Moreover, repeated adjustments of the bound may drastically slow-down
the procedure, forcing it to visit useless tree sub-sections.
For this reason function McSplit was also modified using a top-down
strategy (similar to the one adopted by
k$\downarrow$~\cite{hoffmann2017clique})
by calling the main solve function once per goal size, that is,
$|V_G|$, $|V_G|-1$, $|V_G|-2$, etc., being $G$ the smaller graph
between $G$ and $H$.
This procedure has been proved to beat the original ones in those
cases in which the MCS covers nearly all of the smaller graph.

\begin{figure}[h]
\centering
\begin{subfigure}[b]{0.35\textwidth}
  \includegraphics[width=1.0\textwidth]{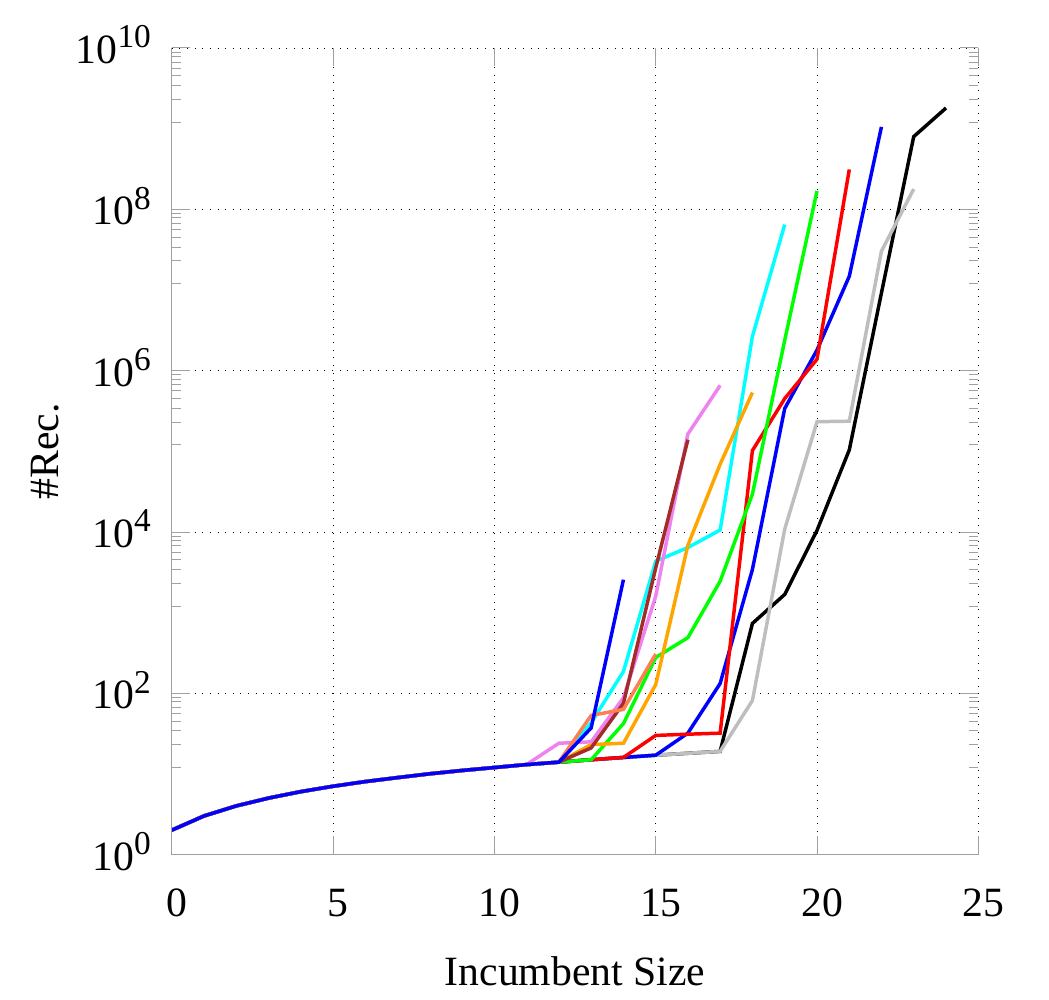}
  \caption{
    \label{plot:partialA}
  }
\end{subfigure}
\begin{subfigure}[b]{0.35\textwidth}
  \label{plot:partial}
  \includegraphics[width=1.0\textwidth]{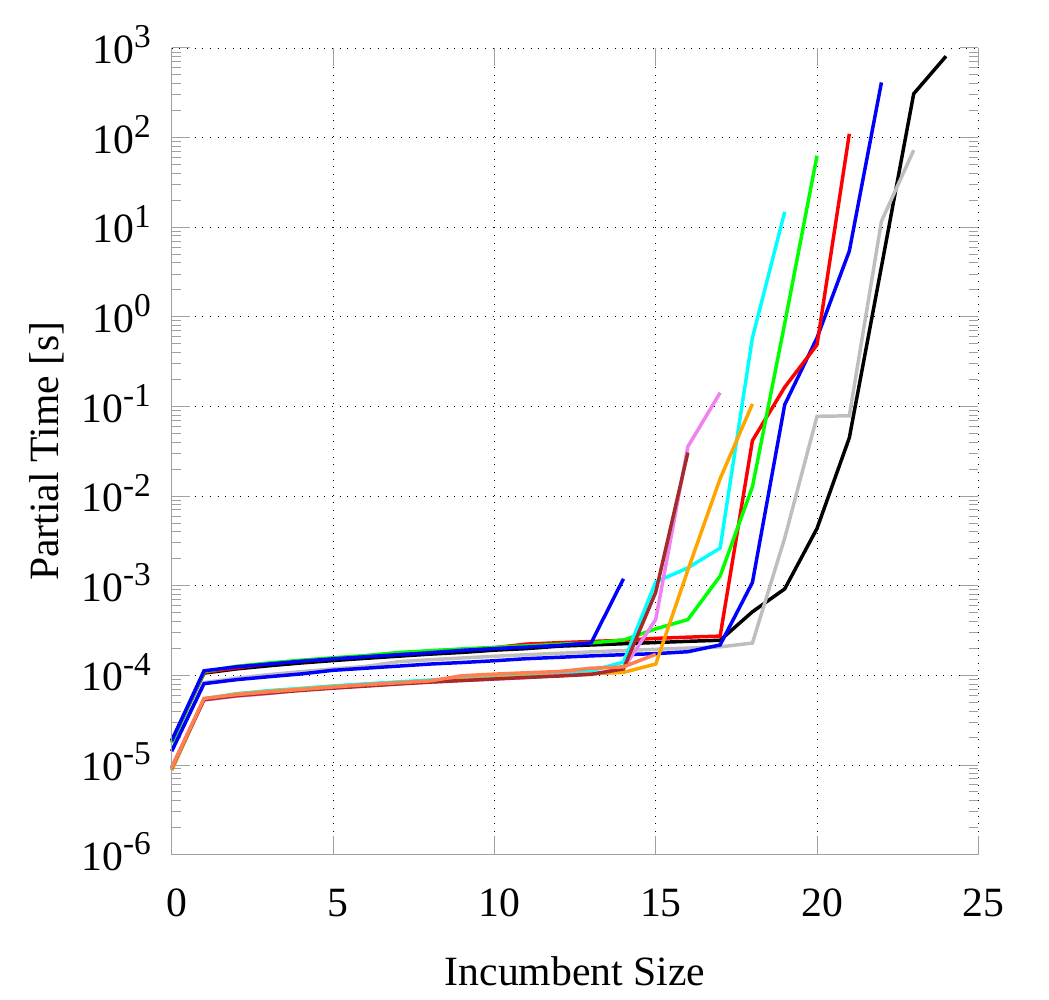}
  \caption{
    \label{plot:partialB}
  }
\end{subfigure}
\caption{
  \label{plot:partial}
  Number of recursions (a) and wall-clock times (b), in the standard
  sequential MsSplit procedure, as a function of the size of the MCS
  discovered.
}
\end{figure}

Moving in this direction, Fig.~\ref{plot:partial} shows the cost of
procedure $mc1$ (the original sequential version of McSplit)
as a function of size of the computed MCS.
For the sake of simplicity, we consider only one representative graph
pair with increasing size of the MCS, size varying from $15$ to $25$. 
Essentially, the two plots report the number of recursions
(Fig.~\ref{plot:partialA}) and the wall-clock times
(Fig.~\ref{plot:partialB}) required to move from one value of the
bound to the next one.
As we use a logarithmic scale along the y-axis, it is easy to infer
that costs drastically increase with the bound, and get quite high
close to the final bound value.

Keeping this consideration in mind, we define ``dead-ends'' the
condition in which no larger MCS is found, but the branch-and-bound
procedure keeps recurring.
Dead-ends re-computation may prove expensive, and we try to forecast
them and to reduce their impact.
To predict dead-ends, we rely on a global criterion such as the number
of recursions.
When this number exceeds a specific threshold without improving the
size of the minimum common subgraph, we consider the current sub-tree
as leading to a dead-end.
Threshold can be evaluated using absolute values or by
taking into consideration the number of recursions that led to the
last increment of the bound.
Once a dead-end is forecast, we experimented with two different
alternatives to reduce its impact.

The first one consists in increasing the bound either by a
little value (usually $1$), or by doubling it.
After varying the bound the procedure may prove that the new bound is
till reachable, finding a higher bound, or unreachable, proving unable
to find a higher bound.
In the first case, we have just avoided several useless iterations,
pruning the search tree much faster and increasing the convergence
speed.
In the second one, we have found an upper bound for the size of the
MCS.
In this case, we look for the correct bound exploiting a binary search
within the lower and the upper bounds.
This search converges in a logarithmic number of recursive checks to
the exact bound.

A second possibility consists in adopting a ``restart'' strategy.
Restart strategies have been proposed for satisfiability (SAT) solvers,
and they have been used for over a decade~\cite{1998Gomes}.
With a restart a SAT solver is forced to backtrack according to some
criterion.
Although, this may not be considered as a sophisticated technique,
there is mounting evidence that it has crucial impact on
performance.
In our case, restarting can be thought of as attempt to avoid spending
too much time in branches in which there is not an MCS larger than the
current one.
We adopt a simple restart strategy for our MCS tools, consisting in
the random selection of the another tree branch within the current
bidomain representation, from which to move on using recursion.
At the same time, to make sure that the branch-and-bound procedures
stop searching only once the entire tree has been visited, for each
restart we store global information to indicate which part of the tree
has already been visited.
This information consists of task object pairs.
Task objects have been introduced in Section~\ref{sec:cpu}, and each
pair indicates the range of recursions already taken into
consideration.
Notice that, to avoid too large data structure and to be efficient,
restarts should not be applied too frequently.
As a consequence, in this case we often trigger a new restart when the
number of recursions has doubled with respect to the last one in which
we found a larger MCS.


In our implementation, all previous engines are orchestrated by a
a Python interface.
This interface may run in two modalities.
In the first one, a number of processes, each one running a different
algorithm or a different configurations of the same algorithms, are run
in parallel.
Whenever a process finishes with a result all others are terminated.
In the second one, we first run only sequential versions for a few
seconds (usually from 5 to 10).
In case a results is not returned, we concurrently run the parallel
CPU and massively parallel GPU versions.
In this last case, the CPU and the GPU versions proceed independently,
even if some information (e.g., the current bound) could be maintained
and updated globally.
The portfolio thus include several tool versions written in C, C++, and
CUDA, plus an high-level script written in Python.
We will refer to the portfolio version as version $vP$ in the sequel.


\section{Experimental Results}
\label{sec:expRes}

In this section we report results for all versions described in
Section~\ref{sec:portfolio} (with and without the modifications
reported in Section~\ref{sec:portfolioModification}), plus the
multi-engine (portfolio) application ($vP$).

Tests have been performed on a machine initially configured for gaming
purposes and equipped with a CPU Intel i7 4790k over-clocked to
4.4~GHz; 16~GiB of RAM at $2,133$~MHz; a GPU NVIDIA GTX 980
over-clocked to 1,300~MHz, with 4~GiB of dedicated fast memory and
2,048 CUDA cores belonging to Compute Level 5.2.
The CPU has 4 physical cores, which benefit from Intel's
Hyper-Treading and are split into 8 logical cores, theoretically
improving multitasking.
The number of threads present in the pool for the parallel CPU version
is set to the number of logical threads that can run in parallel on
the processor, i.e., 8.
The operating system used was Ubuntu 18.04 LTS.


\subsection{Data-sets description}
\label{sec:expResDataSet}

We tested our code using the ARG database of graphs provided by Foggia et al.~\cite{foggia2001database} and De Santo et al.~\cite{DeSanto2003}. The ARG database is composed by several classes of graphs, randomly generated according to six different generation strategies with various parameters settings. The result is a huge data-set of 168 different types of graphs and a total of $166.000$ different graphs\footnote{For more information on the ARG database visit \url{https://mivia.unisa.it/datasets/graph-database/arg-database/}.}.

For our purposes, we selected only a subsection of the data-set, containing graphs pairs already prepared to have maximum common subgraphs of specific dimension. In particular we used the \texttt{mcs10}, \texttt{mcs30}, \texttt{mcs50}, \texttt{mcs70} and \texttt{mcs90} categories, i.e., graphs pairs with maximum common subgraphs corresponding to 10, 30, 50, 70 and 90\% of the original graphs size. For each of these categories several generation strategies are present, such as \textit{bound-valence} graphs (bvg) or graphs generated using 2D, 3D or 4D meshes with different parameters.


Starting from this subset of the ARG database we considered two sub-categories of graphs pairs: Small ones (with less than 40 vertices, 2,750 pairs), and medium (from 40 to 50 vertices, 240 pairs). Graphs larger than 50 vertex are not considered as it has been impossible to solve any problem in less than 1,000 seconds. All graphs are manipulated as undirected and unlabelled to generate the following results.


\subsection{Analysis of the Stand-Alone Procedures}
\label{sec:expResStandAone}

\begin{figure}[h]
\centering
\includegraphics[width=0.35\textwidth]{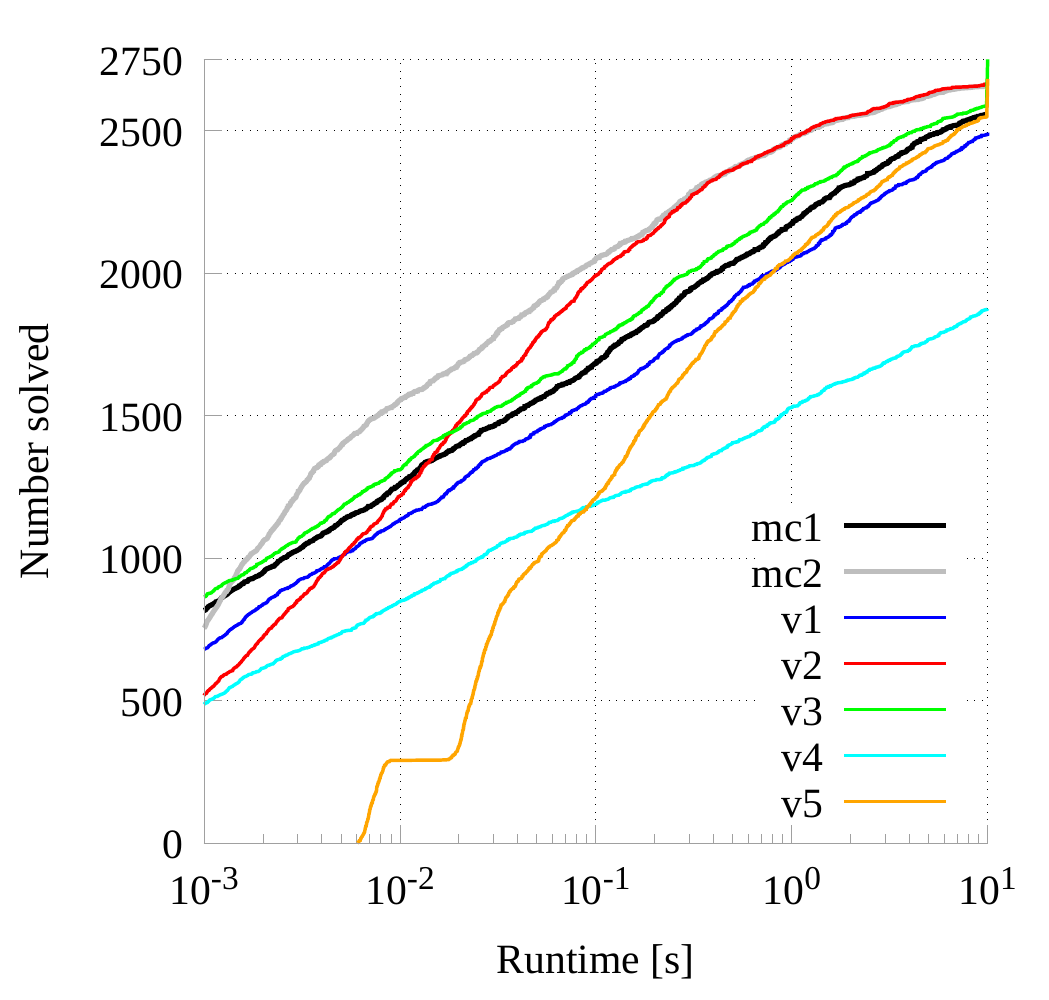}
\caption{
  \label{plot:all}
  Cumulative number of instances (y axis) solved in under a certain
  time (x axis).
}
\end{figure}

Fig.~\ref{plot:all} plots all tested versions on the small graph set with a tight wall-clock time limit, equal to 10 seconds. Given an execution time threshold on the x-axis, the relative y value indicates the number of problem instances solved by the algorithm under such a threshold. The x-axis has a logarithmic scale, while the y-axis is standard linear. For instance, from the $mc2$ plot, we can read that the parallel C++ version solved just over 2,000 instances (out of 2,750) in less than $10^{-1}$ seconds. The sequential version took about four times the time to solve the same number of instances.

Not surprisingly, multi-threaded versions are usually less effective  for very small problems. This is mainly due to the higher computing overhead for threads instantiation. As a consequence, none of the multi-thread versions can be considered faster than any sequential version to solve small and easy instances. Nevertheless, starting from run-times around $10^{-3} \div 10^{-2}$ seconds, parallels version quickly recover the performance gap and actually start performing significantly better for every remaining instances, with speedups up to almost an order of magnitude.

It is also clear that the sequential version ($v3$) performs more
or less as the corresponding recursive version.
On the contrary, the multi-thread iterative version ($v4$) is actually
the worst performing version of the group.
This fact can be explained by remembering that such version was only
developed to explore the possibility to implement the sequential
algorithm in a multi-thread but non-collaborative way.
Thus threads in this implementation are independent, and performances
can better scale by increasing the number of threads, what is
desirable attempting a CUDA implementation.

As expected, due to the high computing overhead necessary to initialize the environment, the CUDA version is significantly slower than any other to start. You can also clearly see that the plot has a flat step in the bottom part. This phenomena is due to the fact that the algorithm runs on the CPU up to the fifth depth level of the virtual research tree and only after such level branches are delegated to GPU threads. In any instance for which the solution turns out to be very small (i.e., a maximum common subgraph with up to 5 vertices), actually the GPU is never called into play, and the computation is entirely done by the CPU. Such instances, in the small set database are about $260$, and they are the first growing part of the plot, up to the step. Then when instances begin to have larger solutions, and thus also the GPU is used to solve them, a gap is formed due to the need of copying data to and from the GPU. For this reason instances that make use of the GPU can not run for less than about $20$ milliseconds, and this determine the gap in the graphic. From there on, performances increases much faster than any other version, actually going to almost equalize CPU parallel versions in number of timeout instances, even with a much slower start.

For the smaller set parallel CPU versions (both the original one in
C++ and our C implementation) are the best performing implementations.
The CUDA implementation, even if not competitive for very easy
instances, is faster then the C sequential versions for run-times
higher than one second.
Moreover, for the harder instances, CUDA is able to close the gap with
the faster implementations, and it can be noted that the tangent line
to the curves in the timeout point, at 10 seconds, is almost
horizontal for the original parallel versions, while it is more
inclined for the CUDA one.
This let us imagine that for harder instances, i.e., requiring more
than $100$ seconds, the CUDA implementation could become faster than
all other implementations.

\begin{figure}[h]
\centering
\begin{subfigure}[b]{0.35\textwidth}
  \includegraphics[width=1.0\textwidth]{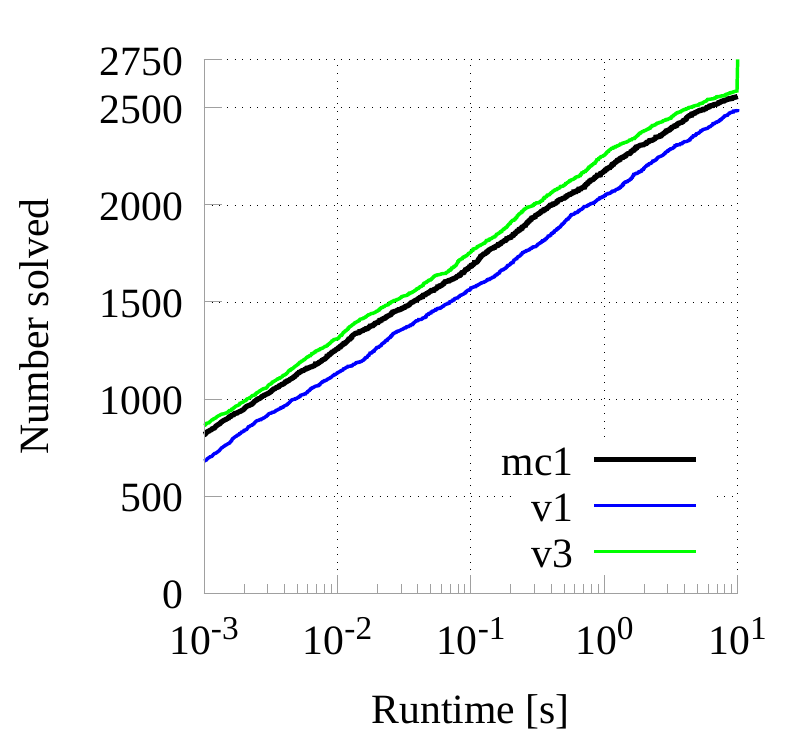}
  \caption{
    \label{plot:seq}
    Sequential summary.
  }
\end{subfigure}
\begin{subfigure}[b]{0.35\textwidth}
  \includegraphics[width=1.0\textwidth]{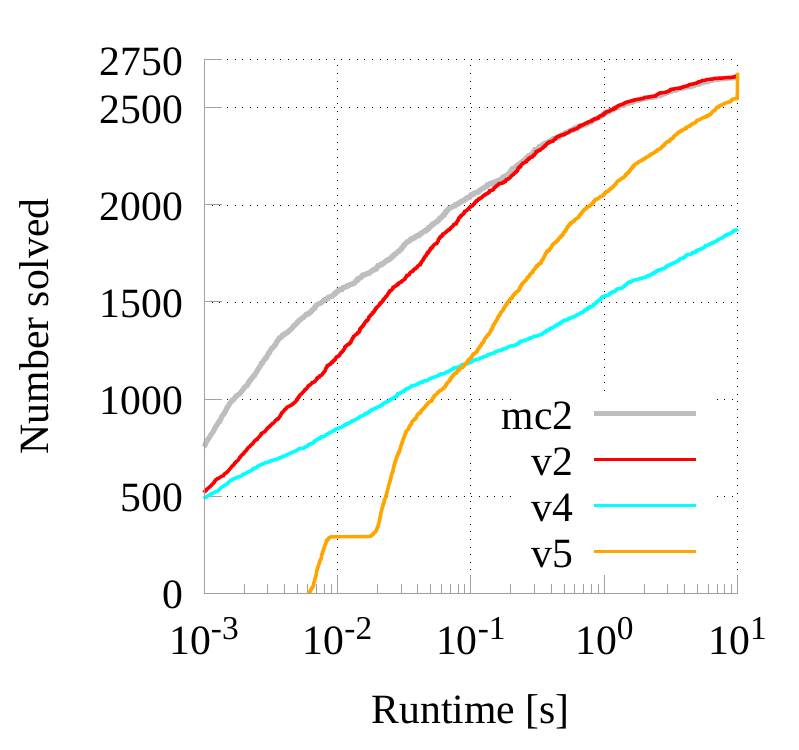}
  \caption{
    \label{plot:par}
    Multi-thread summary.
  }
\end{subfigure}
\caption{
  \label{plot:parAseq}
  Cumulative number of instances (y-axis) solved in under a certain
  time (x-axis) plotted separately for the sequential and parallel
  versions.
}
\end{figure}

Fig.~\ref{plot:parAseq} plots separately sequential and parallel versions.
From Fig.~\ref{plot:seq} it is possible to see that all sequential
versions perform more or less in the same way, with a gap of about
half an order of magnitude between the slowest and the fastest
version.
It is interesting to notice that for the sequential approach, the
iterative approach is the fastest one.
A different situation is shown in Fig.~\ref{plot:par} where the
iterative multi-thread version performs up to two orders of magnitude
worse than the original parallel versions. 
Moreover, it is possible to notice that the parallel C++ version
performs generally better for instances of simpler and medium
difficulty.

\begin{figure}[h]
\centering
\includegraphics[width=0.40\textwidth]{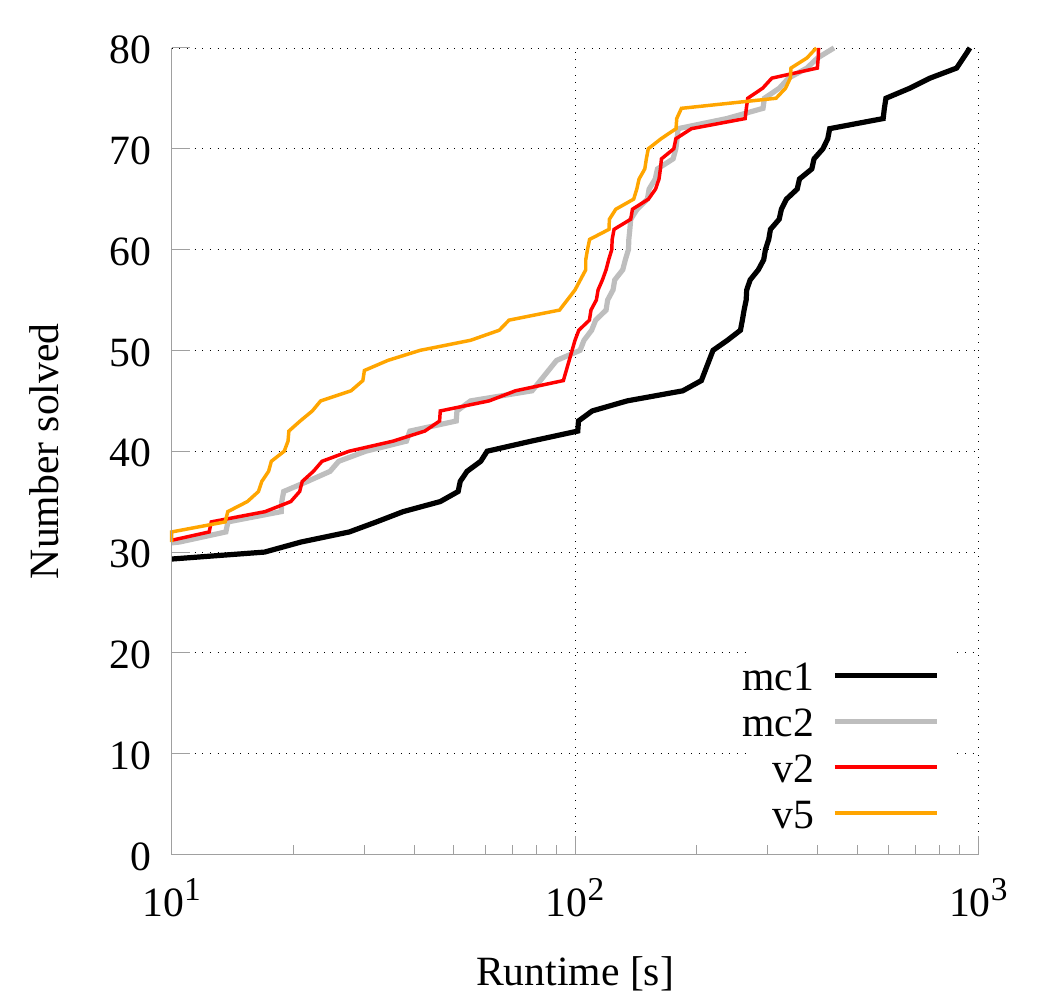}
\caption{
  \label{plot:medium}
  Cumulative number of instances solved in a given runtime by the fastest
  CPU versions ($mc1$, $mc2$ and $v2$) and by the GPU ($v5$) implementation.
}
\end{figure}

Fig.~\ref{plot:medium} reports the plots for the small and
the medium graph sets considered together.
The allowed wall-clock time has been increase to $1000~s$.
In this picture, we just consider the fastest strategies of
Fig.~\ref{plot:all}.
As it can be noticed, as previously imagined, the CUDA version closes
the gap with the parallel CPU implementation around $20$ seconds, and
it is definitively faster than the CPU version up to about $400$
seconds.

\begin{figure}[h!]
\centering
\includegraphics[width=0.48\textwidth]{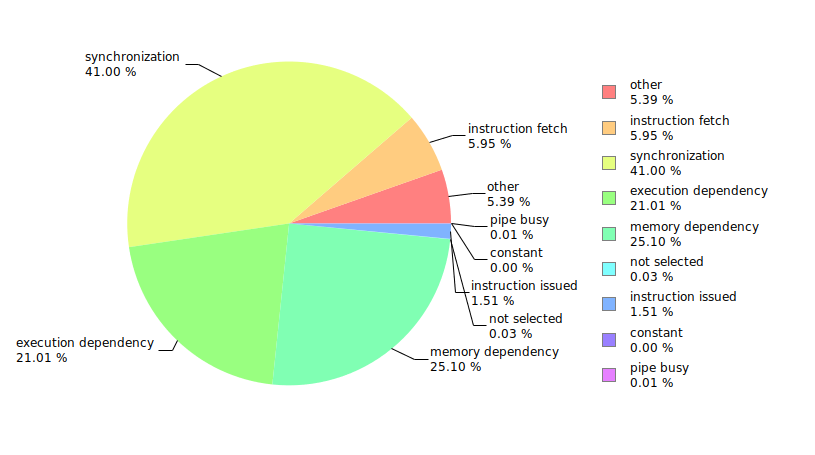}
\caption{
  \label{fig:stall_chart}
  The chart represents all main reasons forcing the CUDA kernel
  function to stall, decreasing performances.
}
\end{figure}

Two main aspects influence the performances of our CUDA
implementation.
They are the \textit{principle of data locality} and the
\textit{branch divergence}.
The first one needs to be maximized, so that when threads in the same
warp access a variable in their local memory, all requests can be
grouped in one single memory fetch of consecutive memory locations.
In this way, the kernel better use the memory bandwidth.
The second aspect, needs to be minimized.
In fact, when threads in the same warp hit a branch instruction
and they take different routes, the whole warp execute both the routes
sequentially, with part of the treads stalling for the first route
and part of the threads stalling for the second ones.
In this way the different branch paths are not executed in parallel
but sequentially, and their execution time is added up.
Moreover, if one or a few threads take a significantly longer route,
compared to any other thread, during the same kernel execution, it or
they will necessarily be waited by every other thread.
Fig.~\ref{fig:stall_chart} represents an analysis of the main
reasons forcing our working threads in a stall situation.

Three main reasons can be easily identified: Synchronization, memory
dependency, and execution dependency.
Synchronization issues, with $41$\% of the overall stalls, happen when
a thread is blocked on a \texttt{\_\_syncthreads} instruction.
Unfortunately, for the \textsc{McSplit} algorithm the workload for
different threads is inherently heavily unbalanced.
Thus many of them stall at the end of their execution, just waiting
that slower threads finish their work before finding the current best
solution found during the kernel execution.
The second main reason for stalling, i.e., the memory dependencies, is
due to those cases in which a load or store operation needs to wait
for free memory bandwidth.
This situation happens quite frequently in our code, where the data
locality principle is limited, because each thread has a separate
stack to simulate recursion, and the stack is allocated in the local
memory of each thread.
Thus, every time a location of the stack is accessed, a memory fetch
has to be performed, and such requests cannot be coalesced within
threads in the same warp, because each threads access to different
cells of respective stack that are not close in memory.
The third significant for thread stalling, i.e., execution dependency,
happens when an instruction is stalled waiting for one or more
arguments to be ready.
This problem can be avoided by increasing instruction-level
parallelism, but of course improving it is not that simple, because we
also have to take into account the number of registers reserved for
each thread, and the one available for each block.
This is an important limitation to increase the number of threads and
the instruction parallelism.

\begin{figure}
\centering
\includegraphics[width=0.48\textwidth]{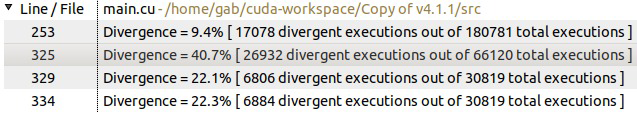}
\caption{
  \label{fig:divergence}
  Extract from the profiling analysis executed by Nsight.
  The analysis shows the most divergent instructions in our CUDA main
  code section.
}
\end{figure}

In addition to the stalls chart, it is also interesting to observe
that some branch instructions in the kernel function show an high
amount of divergence during the execution.
Fig.~\ref{fig:divergence} indicates the most divergent
instructions inside the body of the kernel function discovered by
Nsight\footnote{NVIDIA Nsight is a debugger, analysis and profiling
tool for CUDA GPU computing (\url{https://developer.nvidia.com/tools-overview}).}.
Note that the worst one, the one at line $325$, corresponds to the
branch where the bound of the current branch is computed and the
branch is eventually pruned.
When the tree branch is not pruned, the body of the main loop of
the kernel function is executed to compute new label domains and new
solutions.
On the contrary, when the branch is pruned, the code jump to the next
main iteration.
The overall behavior is then significantly different.
All other relevant branches with high divergence (around $20$\%), are
again inside the main loop of the kernel function.
This lead to overall high percentage of branch divergence in the core
of the kernel function, limiting a lot the maximum instruction
throughput reachable by the kernel.

\subsection{Analysis of the Portfolio Approach}
\label{sec:expResPortfolio}

We insert in our portfolio the versions of the MCS procedure described
in Section~\ref{sec:portfolio}. Versions $v2$, $v3$ and $v5$ have also
been augmented with the heuristics and the settings described in
Section~\ref{sec:portfolioModification}.

\begin{figure*}
\centering
\begin{subfigure}[b]{0.30\textwidth}
  \includegraphics[width=1.0\textwidth]{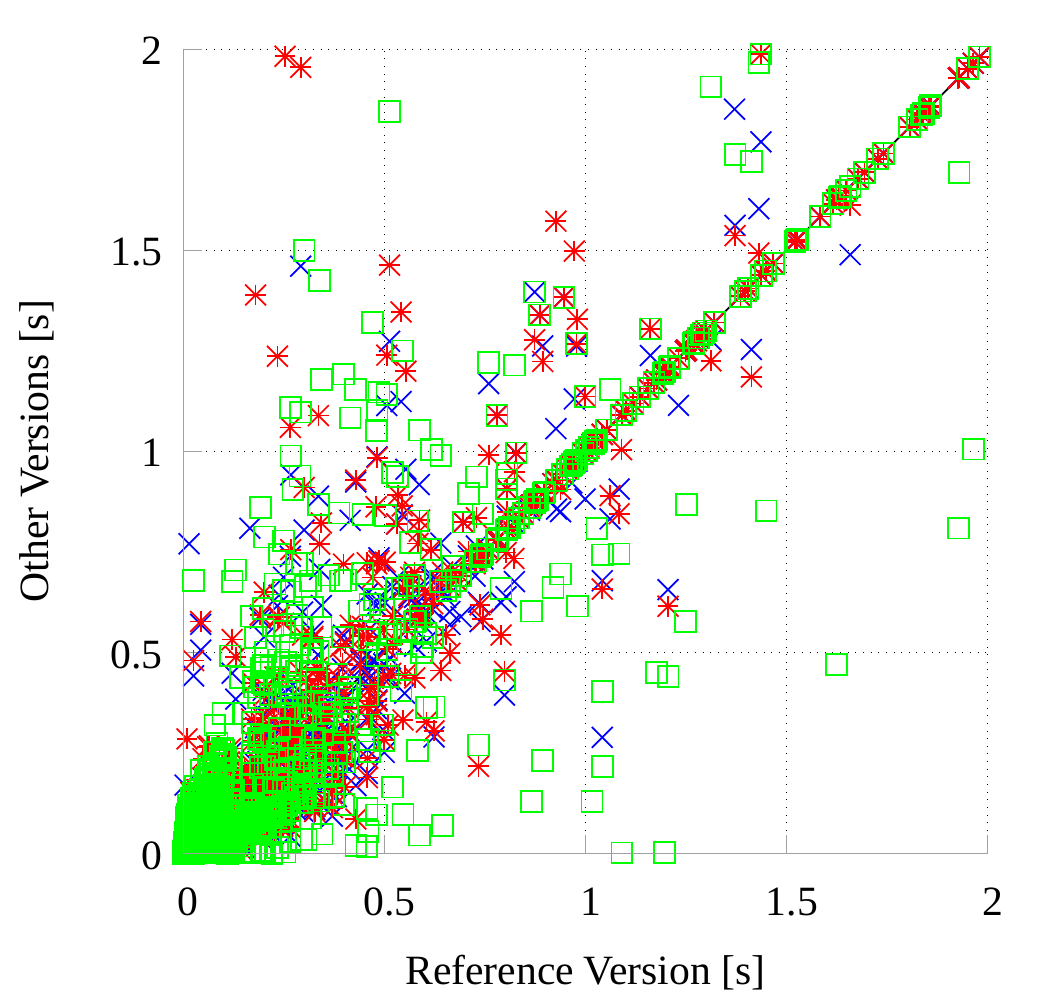}
  \caption{\label{fig:scatterPlot_a}}
\end{subfigure}
\hfill
\begin{subfigure}[b]{0.30\textwidth}
  \includegraphics[width=1.0\textwidth]{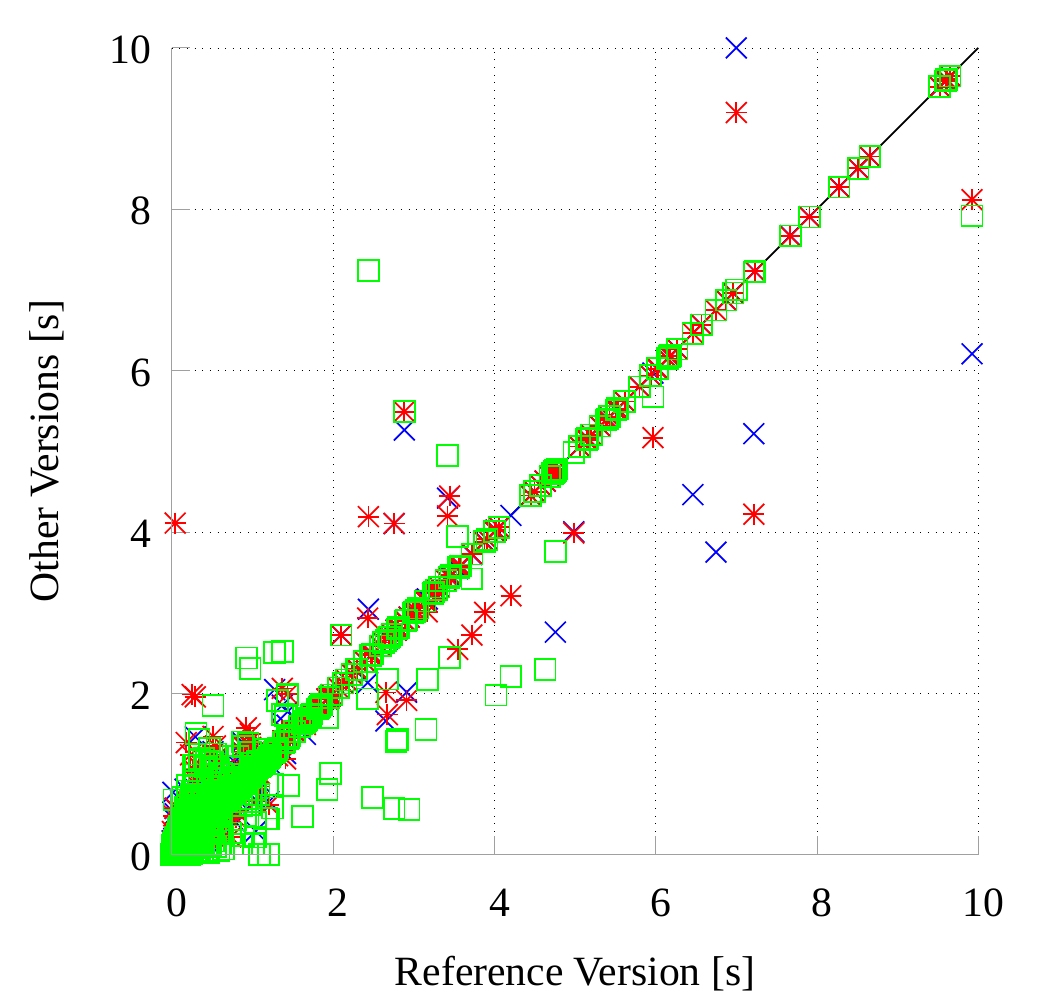}
  \caption{\label{fig:scatterPlot_b}}
\end{subfigure}
\hfill
\begin{subfigure}[b]{0.30\textwidth}
  \includegraphics[width=1.0\textwidth]{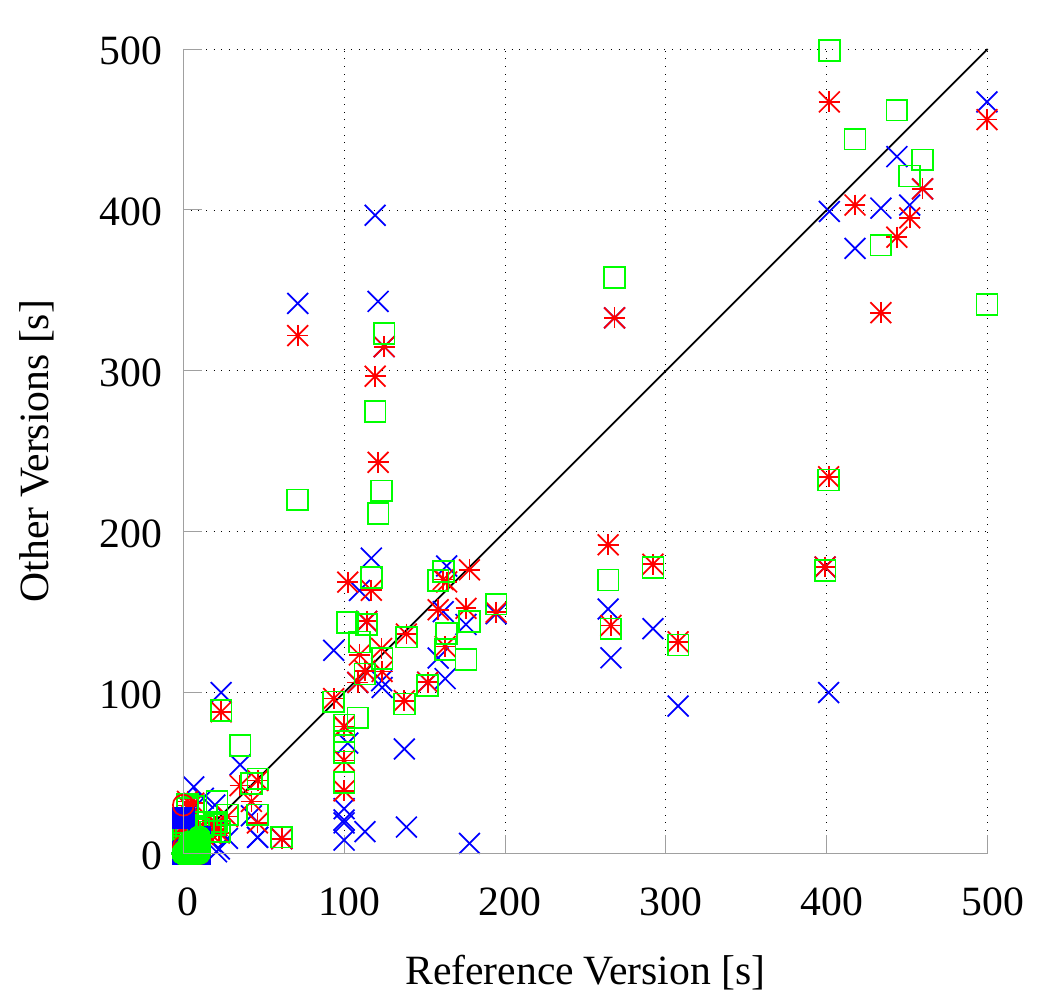}
  \caption{\label{fig:scatterPlot_c}}
\end{subfigure}
\caption{
  \label{fig:scatterPlot}
  Running times for different heuristics (reported on the y-axis)
  implemented within the sequential version (reported on the x-axis)
  denoted as $v3$ and taken as reference version.
  These heuristics include the reordering strategy for the adjacency
  matrix (blue crosses), the dead-end prediction with bound correction
  (red stars), and the dead-end prediction with randomize restart
  (green boxes).
  Plots~\ref{fig:scatterPlot_a},~\ref{fig:scatterPlot_b},
  and~\ref{fig:scatterPlot_c} focus on different time ranges, i.e.,
  0--2, 0--10, 0-500 seconds.
}
\end{figure*}

To analyze the potential benefit of our portfolio approach,
Fig.~\ref{fig:scatterPlot} compares the performance of the original
version $v3$ modified with 3 different heuristics.
These settings include our reordering strategy for the adjacency
matrix, our dead-end prediction with bound correction, and the
dead-end prediction with randomize restart.
The 3 plots report the running times for the original (reference)
version on the x-axis, whereas they represent with different symbols
(and colors) the running times for the 3 modified versions.
Points on the main diagonal include benchmarks on which the
performance is the same.
Fig.~\ref{fig:scatterPlot_a} focuses on small graph pairs which can be
solved within 2 seconds, Fig.~\ref{fig:scatterPlot_b} concentrates on
a time range up to 10 seconds, and Fig.~\ref{fig:scatterPlot_c}
extends the analysis to 500 seconds. As it can be noticed, different
heuristics perform quite differently on different graph pairs even if
they do not really improve the original version on average.
For example, in the time range fro 0 to 10 seconds, the original
version $v2$ is the fastest method in 34\% of the cases, whereas the 3
heuristics make it faster in 26\%, 23\%, and 13\% of the cases,
respectively (all versions run out of time in the remaining 4--5\% of
the cases).
Similar considerations hold for the procedures described in
Section~\ref{sec:portfolio}.
A more accurate analysis of the the data plotted in
Fig.~\ref{plot:all} show that the number of instances on which the
various methods are the fastest ones are the following: $mc1$ 60,
$mc2$ 1404, $v1$ 10, $v2$ 245, $v3$ 947, $v4$ 0, and $v5$ 4.
The same conclusions can be derived from the cumulative plot if
Fig.~\ref{plot:medium}, as the CPU and the GPU procedures have similar
performance on average, but they present quite different performance
on a benchmark-by-benchmark basis.

\begin{figure}
\centering
\begin{subfigure}[b]{0.35\textwidth}
  \includegraphics[width=1.0\textwidth]{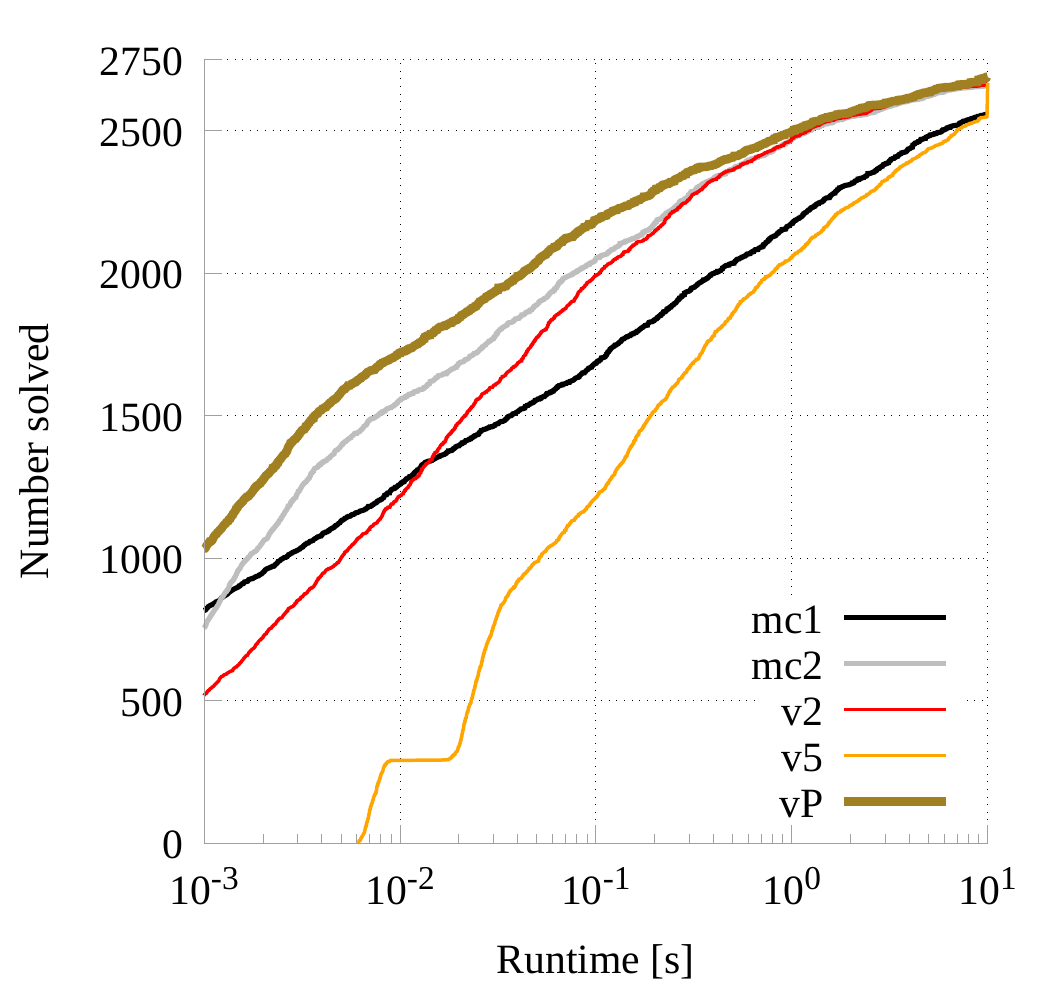}
  \caption{\label{fig:portfolioPlot_a}}
\end{subfigure}
\begin{subfigure}[b]{0.35\textwidth}
\includegraphics[width=1.0\textwidth]{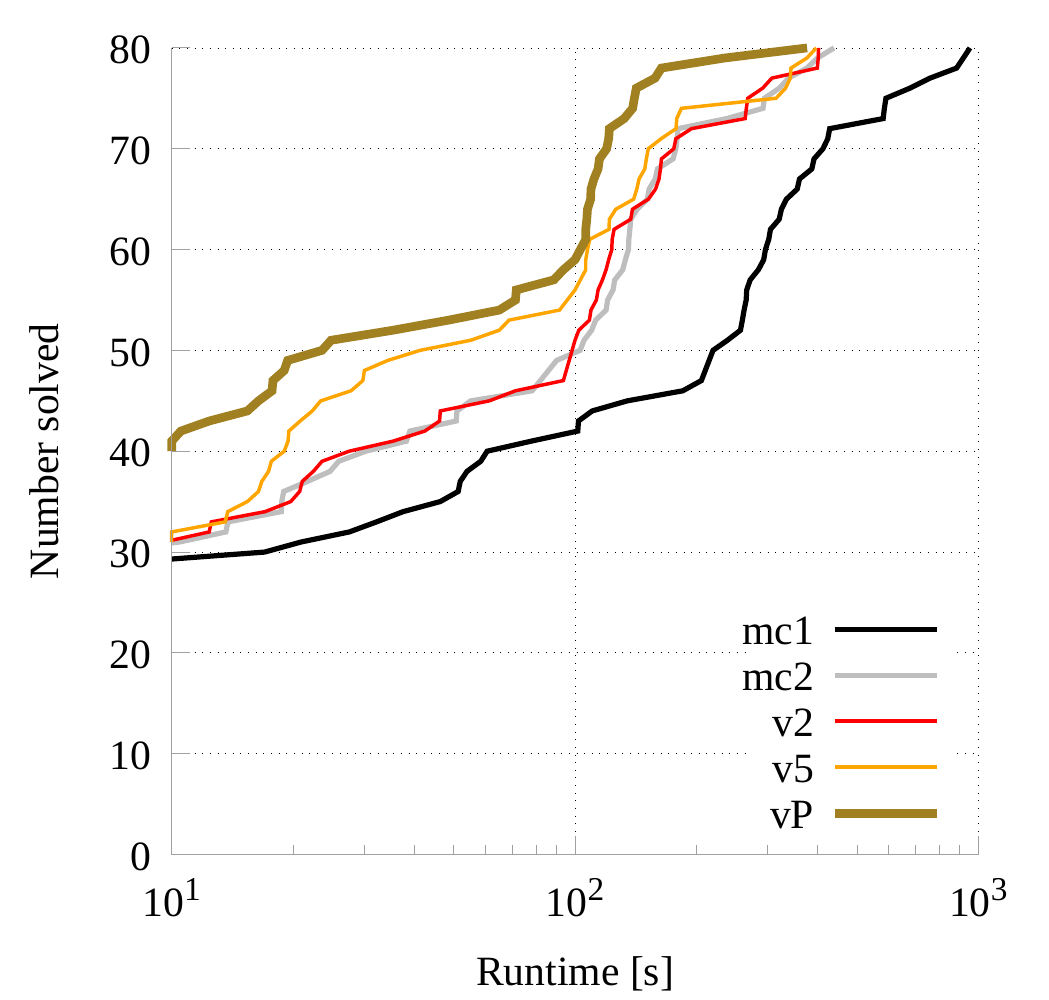}
\caption{\label{fig:portfolioPlot_b}}
\end{subfigure}
\caption{
  \label{fig:portfolioPlot}
  The most efficient sequential and parallel McSplit versions compared
  against the portfolio with a time limit of (a) $10$ seconds and (b)
  $1,000$ seconds.
}
\end{figure}

Results gathered with the portfolio approach are presented in
Fig.~\ref{fig:portfolioPlot}.
The two graphics plot the most efficient approaches (as
represented if Fig.~\ref{plot:all} and Fig.~\ref{plot:medium})
with the one gathered with our portfolio ($vP$).
As in the previous cases, Fig.~\ref{fig:portfolioPlot_a} includes data
for small graphs analyzed up to 10 seconds.
Fig.~\ref{fig:portfolioPlot_b} extends our analysis for medium graphs.
and a time limit up to 1000 seconds.
It is easy to argue that the portfolio is faster on average and able to
solve more instances that any other method on both wall-clock time
(and graph size) ranges.

\section{Conclusion}
\label{sec:conclusion}

In this work, we extend the algorithm named McSplit, a
state-of-the-art branch-and-bound algorithm to find the maximum common
subgraphs between two graphs.

First, we parallelize it, both to as a multi-core CPU implementation and as a CUDA-based GPU many-core one. Concerning the CPU procedure, we experimented with several parallel approaches, going from using different data structures to adopting several divide-and-conquer paradigms, with the purpose of balancing the threads workload and to keep under control the memory usage. As far as the CUDA extension is concerned, the main issue is that, like many other graphs algorithms, the analyzed one is based on complex data structures and on recursion. Both features do not usually suit well with GPU computing, and we propose a few ways to overcome these limitations.

Then, we develop new heuristics to reorder the adjacency matrix, trying to isolate matrix blocks and make them upper triangular, to smooth the heavy-tail phenomenon dealing with dead-ends, and to restart the process with automatic restarts. These algorithms, albeit not more efficient than the original ones from an average point of view, allow a significant performance improvement for specific graph pairs. 

Finally, we propose a portfolio approach, integrating several local search algorithms as component tools. In this portfolio, a python interface orchestrate many CPU versions with one GPU procedure, exploiting modern hardware resources at their best. We demonstrate that the portfolio approaches is beneficial in the domain, and CUDA versions may worth further analysis.


\bibliographystyle{IEEEtran}
\bibliography{./main.bbl}



\begin{IEEEbiography}[{\includegraphics[width=1in,height=1.25in,clip,keepaspectratio]{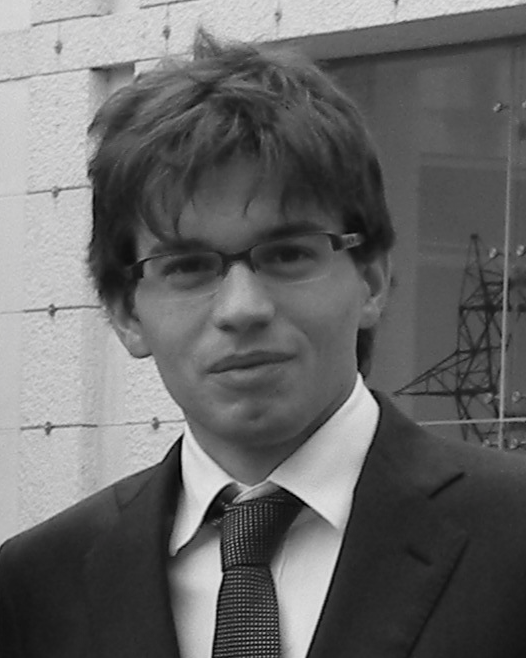}}]{Andrea Marcelli}
received his M.Sc. degree in Computer Engineering from Politecnico of
Torino, Italy, in 2015. Currently he is a Ph.D. student in Computer
and Control Engineering at the same institute and member of the CAD
group. His research interests include malware analysis,
semi-supervised modeling, machine learning and optimization problems,
with main applications in computer security.
\end{IEEEbiography}

\begin{IEEEbiography}[{\includegraphics[width=1in,height=1.25in,clip,keepaspectratio]{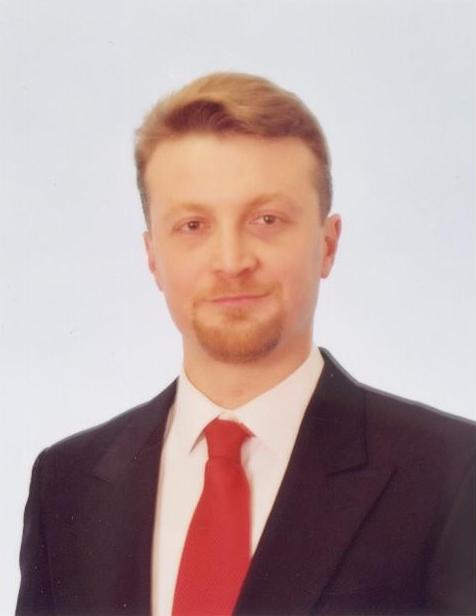}}]{Stefano Quer}
received a M.S. in Electronic Engineering from Politecnico
di Torino in $1991$, and a Ph.D. degree in Computer Engineering, from
the Ministry of University and Scientific and Technological Research in
Rome in $1996$.
He has been a Visiting Faculty in the Department of Electronic Engineering
and Computer Science of the University of California in Berkeley.
He has been an intern with the ``Advanced Technology Group'' at Synopsys Inc.,
Mountain View, California, and with the ``Alpha Development Group''
at Compaq Computer Corporation, Shrewsbury, Massachussetts.
He has been a Compaq Computer Corporation consultant.
He is currently professor with the Department of Control and
Computer Engineering at Politecnico di Torino, Torino, Italy.
His main research interests include systems and tools for CAD for VLSI,
formal methods for hardware and software systems, and embedded
systems.
Other activities focus on the development of sequential and concurrent
algorithms and on optimization techniques able to achieve acceptable
solutions with limited resources.
\end{IEEEbiography}

\begin{IEEEbiography}[{\includegraphics[width=1in,height=1.25in,clip,keepaspectratio]{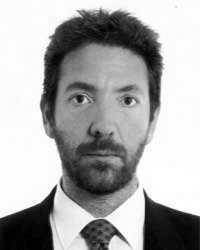}}]{Giovanni Squillero}
(M01-SM14) received his M.S. and Ph.D. in computer engineering in 1996 and 2001, respectively. He is currently an associate professor of computer science at Politecnico di Torino. His research mixes the whole spectrum of bio-inspired metaheuristics with computational intelligence, machine learning, and selected topics in electronic CAD, games, multi-agent systems. Other activities focus on the development of optimization techniques able to achieve acceptable solutions with limited amount of resources, mainly applied to industrial problems. 
\end{IEEEbiography}

\EOD

\end{document}